\documentclass[11pt]{article}

\usepackage{graphicx,epsfig}
\usepackage{multicol}
\usepackage{amsmath,amssymb,cite,color,hyperref}

\newcommand{\sqrts}{\sqrt{s}}
\newcommand{\pp}{$p$-$p$}

\newcommand{\pt}{p_{_{\rm T}}}

\newcommand{\ETg}{E_{_{\rm{T}}}^{\gamma}}

\newcommand{\jetphox}{{\sc jetphox}}
\newcommand{\pythia}{{\sc pythia}}

\def\mean#1{\ensuremath{\left<#1\right>}}

\newcommand{\be}{\begin{equation}}
\newcommand{\ee}{\end{equation}}
\newcommand{\bea}{\begin{eqnarray}}
\newcommand{\eea}{\end{eqnarray}}
\newcommand{\bi}{\begin{itemize}}
\newcommand{\ei}{\end{itemize}}
\newcommand{\ben}{\begin{enumerate}}
\newcommand{\een}{\end{enumerate}}
\newcommand{\la}{\left\langle}
\newcommand{\ra}{\right\rangle}

\newcommand{\lp}{\left(}
\newcommand{\rp}{\right)}

\def\frac#1#2{{{#1}\over {#2}}}
\def\gsim{\mathrel{\rlap{\lower4pt\hbox{\hskip1pt$\sim$}}
    \raise1pt\hbox{$>$}}}         %greater than or approx. symbol
\def\lsim{\mathrel{\rlap{\lower4pt\hbox{\hskip1pt$\sim$}}
    \raise1pt\hbox{$<$}}}         %less than or approx. symbol

\newcommand{\draft}[1]{}

\definecolor{grey}{rgb}{0.5,0.5,0.5}

\begin{document}
\hfill {\sf CERN-PH-TH/2013-006}
%\begin{flushright}
%CERN-PH-TH/2011-302\\
%\end{flushright}
\vspace{1.cm}

\begin{center}
{\Large\bf Sensitivity of the LHC isolated-$\gamma$+jet data to the\\\vspace{0.2cm} parton distribution functions of the proton}
\end{center}
\vspace{0.5cm}

\begin{center}
L.~Carminati$^{1,2}$, G. Costa$^{1}$, D.~d'Enterria$^{3}$, I.~Koletsou$^{1}$,\\
G. Marchiori$^{4}$, J.~Rojo$^{5}$, M.~Stockton$^{6}$, F. Tartarelli$^{1}$
\vspace{0.5cm}

{\it $^1$ INFN Sezione di Milano, Milano, Italy\\}
{\it $^2$ Dpt Fisica, Universit\`a di Milano, Milano, Italy\\}
{\it $^3$ CERN, PH Department, CH-1211 Geneva 23, Switzerland\\}
{\it $^4$ LPNHE, Univ. Pierre et Marie Curie - Univ. Paris-Diderot - CNRS/IN2P3, Paris, France\\}
{\it $^{5}$ CERN, PH Department, TH Unit, CH-1211 Geneva 23, Switzerland\\}
{\it $^{6}$ Dept of Physics, McGill University, Montreal, Quebec - Canada}
\end{center}

\vspace{0.4cm}

\begin{center}
{\bf \large Abstract}
\end{center}
%In this note 
We study the impact of differential isolated-photon+jet cross sections measured in proton-proton collisions 
at a center-of-mass energy of $\sqrt{s}$ = 7~TeV on the parton distribution functions (PDF) of the proton.
Next-to-leading-order perturbative QCD (pQCD) calculations complemented with the NNPDF2.1 parton densities, 
and a Bayesian PDF reweighting method are employed. We find that although the current data provide 
only mild constraints to the parton densities, future $\gamma$-jet measurements with reduced
experimental uncertainties can improve our knowledge of the gluon density over a wide 
range of parton fractional momenta $x$ as well as of the quarks at low-$x$. 

%\clearpage

%\tableofcontents

%%%%%%%%%%%%%%%%%%%%%%%%%%%%%%%%%%%%%%%%%%%%%%%%%%%%%%%%%%%%%%%%%%%%%%%%%%%%%%%%%%%%%%%%

%\scrollmode

\section{Introduction}

The accurate determination of the parton distribution functions (PDF) of the proton in a wide range of
momentum fractions $x$ and energy scales $Q$~\cite{Perez:2012um} is a crucial ingredient for precision studies of the 
Standard Model and new physics at the Large Hadron Collider (LHC)~\cite{Watt:2011kp,Ball:2012wy,Forte:2010dt}.
The availability of new precision data from the LHC covering a large ($x,Q^2$) range -- including processes such
as gauge boson production in association with jets and heavy quarks which hitherto have not been used for PDF
determinations -- provides significant improvements in the accuracy of global PDF fits~\cite{Ball:2012cx}.
Among the processes available in proton-proton (\pp) collisions at the LHC, inclusive prompt-$\gamma$
production -- defined as the production of photons not issuing from the electromagnetic decays of hadrons --
proceeds through the dominant quark-gluon ``Compton'' process $qg\rightarrow \gamma q$
and has been shown to provide direct quantitative constraints on the gluon density
$g(x,Q^2)$~\cite{d'Enterria:2012yj}. In this paper we revisit the phenomenological study carried out 
in~\cite{d'Enterria:2012yj} for inclusive isolated-$\gamma$ 
spectra~\cite{Khachatryan:2010fm,Aad:2010sp,Chatrchyan:2011ue,Chatrchyan:2012vq,arXiv:1108.0253}, 
but focusing now on the newly available $\gamma$-jet data
from the LHC~\cite{ATLAS:2012ar} whose constrained kinematics given, at leading order, by the concurrent
measurement of the photon and back-to-back parton, has the potential to provide additional constraints on the proton PDF.
This is, to our knowledge, the first time that isolated-$\gamma$+jet data in high-energy hadronic collisions 
have been used to assess their sensitivity to the proton parton densities.

\section{Experimental data}  

The experimental $\gamma$-jet cross sections studied are those measured by the ATLAS experiment in \pp\ at 7~TeV  
in a data-sample corresponding to $\sim 37$~pb$^{-1}$ integrated luminosity~\cite{ATLAS:2012ar}.
Isolated photons have been reconstructed in the rapidity range $|y^\gamma|<1.37$ with a transverse energy
$\ETg > 25$~GeV, requiring a total transverse energy below 4~GeV inside a cone of radius $\Delta R=0.4$ 
in pseudorapidity-azimuth %($\eta\times\phi$) 
along the photon direction. 
Jets have been reconstructed with the anti-$k_{T}$ algorithm~\cite{Cacciari:2008gp} with radius parameter $R = 0.4$,
within $|y^{\rm jet}|<4.4$ and for transverse momenta $\pt^{\rm jet} > 20$~GeV. 
The differential cross sections $d\sigma/d\ETg$ are then measured as a function of the photon transverse
energy in six jet-photon angular configurations: $|y^{\rm jet}| < 1.2$, $1.2 \leq |y^{\rm jet}| < 2.8$ and 
$2.8 \leq |y^{\rm jet}| < 4.4$, for the same $(y^\gamma y^{\rm jet}\ge 0)$ and opposite
$(y^\gamma y^{\rm jet}< 0)$ hemispheres. 

%%%%%%%%%%%%%%%%%%%%%%%%%%%%%%%%%%%%%%%%%%%%%%%%%%%%%%%%%%%%%%%%%%%%%%%%%%%%%%%%%%%%%%%%
\section{Theoretical setup}
%\label{sec:th}

The theoretical $\gamma$-jet cross sections have been computed at next-to-leading-order
(NLO) accuracy with the \jetphox\ (version 1.3.0) Monte Carlo (MC) code~\cite{jetphox,Aurenche:1992yc,Catani:2002ny} 
complemented with 100 replicas of the NNPDF2.1 parton densities\footnote{Use of the more recent NNPDF2.3
set~\cite{Ball:2012cx}, which includes LHC data but which came available only after this analysis was
performed, is expected to give the same conclusions as our study.}~\cite{Ball:2011mu,Ball:2011uy}.
The default renormalisation, factorisation and fragmentation scales 
are all set equal to the photon transverse energy, $\mu_{_{\rm R}}=\mu_{_{\rm F}}=\mu_{_{\rm ff}}=\ETg$.
The parton--to--photon fragmentation functions used are the BFG-II (``large gluon'')
set~\cite{Bourhis:1997yu}. The MC photon isolation and jet reconstruction criteria are matched as
closely as possible to each of the experimental cuts. The $\gamma$-jet distributions have been corrected
with \pythia~\cite{Sjostrand:2006za} for non-perturbative effects due to hadronization and \pp\ underlying event.\\

In order to quantify the impact of the photon-jet cross sections on the PDF, we use the Bayesian reweighting
method described in Refs.~\cite{Ball:2010gb,Ball:2011gg} (the same technique could have been performed using
PDF sets based on Hessian error matrices, such as CT10~\cite{Lai:2010vv} and/or MSTW08~\cite{Martin:2009iq},
as discussed in~\cite{Watt:2012tq}). We compare each of the experimental distributions (with $N_{\rm dat}$
data points) to the theoretical prediction obtained with each of the $k$~=~1,..., $N_{\rm rep}$=100 replicas
of the NNPDF2.1 set, defining a data--theory goodness-of-fit $\chi^{2}_k$ as
\be
\chi^{2}_k = \frac{1}{N_{\rm dat}}\sum_{i=1}^{N_{\rm dat}}
\frac{\lp \sigma_i^{{\rm (th)},(k)}-\sigma_i^{{\rm (exp)}}\rp^2}{\Delta_{\rm tot}^2} \ ,
\label{eq:chi2} 
\ee
where $\sigma_i^{{\rm (th)},(k)}$ is the NLO theoretical prediction for the photon-jet cross section obtained
with the $f_k$ PDF replica, $\sigma_i^{{\rm (exp)}}$ is the corresponding experimental measurement, and 
$\Delta_{\rm tot}$ accounts for the experimental statistical and systematic uncertainties added in
quadrature\footnote{The full experimental covariance matrix is not available for this measurement -- the
usefulness of future $\gamma$-jet data would be increased if this covariance matrix is provided (see
discussion in Section~\ref{subsec:pseudodata}).}.
Theoretical uncertainties due to scale variations are between 15\% (low $\ETg$) and 10\% (high $\ETg$) -- as
obtained from the envelope of the theoretical spectra obtained varying the 3 scales in the following 6 combinations: 
($\mu_{_{\rm R}},\mu_{_{\rm F}},\mu_{_{\rm ff}}$)/$\ETg$ = (1/2,1,1) (2,1,1) (1,1/2,1) (1,2,1) (1,1,1/2) (1,1,2)
(1/2,1/2,1/2) (2,2,2) -- but have not been included in the $\chi^2$ analysis as there is not yet a recipe
to consistently include scale uncertainties in global PDF analysis.
The corresponding weight of each replica is then obtained following~\cite{Ball:2010gb,Ball:2011gg}.
These weights $w_k$, divided by the number of MC replicas of the prior PDF set ($N_{\rm rep}$), 
give the probabilities of the replicas $f_k$ given the $\chi_{k}^2$ values to the newly added experimental results.\\

For each set of weights we also compute the rescaling parameter $\alpha$~\cite{Ball:2010gb,Ball:2011gg},
which indicates the value by which one should scale the experimental and/or theoretical
uncertainties in order to achieve a goodness-of-fit $\chi^2_k\approx$~1. 
The distribution of the rescaling variable $\alpha$, normalized to unity,
is used to investigate if any potential disagreement between NLO pQCD and a given dataset
could be due to possibly over-/under-estimated uncertainties, and/or indicate possible
tensions with other datasets.\\

\begin{figure}[htpb!]
\centering
\epsfig{width=0.70\textwidth,figure=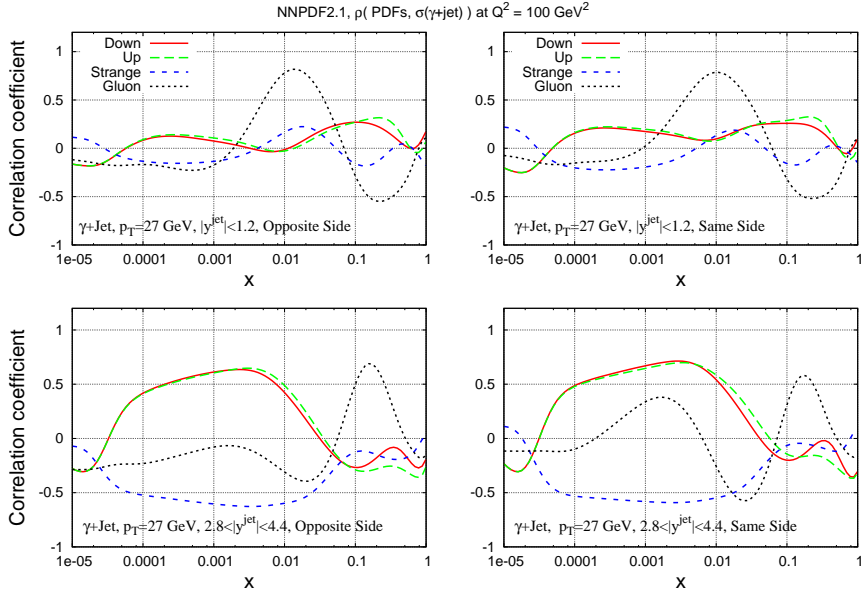}
\caption{\small 
Correlations between the $\gamma$-jet cross section in \pp\ collisions at 7 TeV 
(for the smallest value of the photon transverse energy, $\ETg$~=~27.5~GeV)
and various flavours of the NNPDF2.1 parton densities for central $y^{\rm jet}$ (top) 
and forward $y^{\rm jet}$ (bottom) for the opposite (left) and same-side (right) photon-jet hemispheres. 
\label{fig:correlations1}}
\end{figure}

\begin{figure}[htpb!]
\centering
\epsfig{width=0.70\textwidth,figure=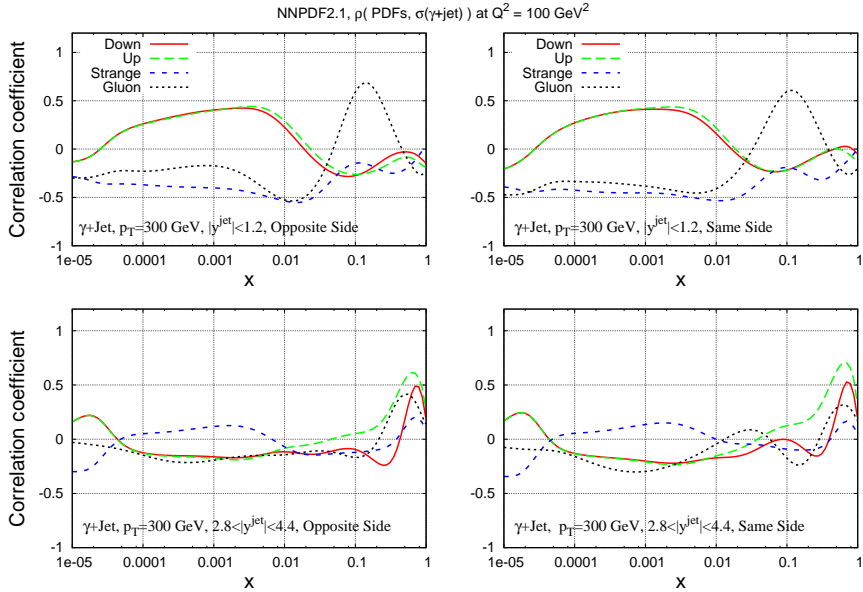}
\caption{\small 
Correlations between the $\gamma$-jet cross section in \pp\ collisions at 7 TeV 
(for the largest values of the photon transverse energy, $\ETg$~=~300~GeV) 
and various flavours of the NNPDF2.1 parton densities for central $y^{\rm jet}$ (top) 
and forward $y^{\rm jet}$ (bottom) for the opposite (left) and same-side (right) photon-jet hemispheres. 
\label{fig:correlations2}}
\end{figure}

The dependence of the measured $\gamma$-jet cross sections on the individual flavour of the
underlying parton densities can be quantified by computing the correlation coefficient between each of the
light-quark and gluon distributions and the NLO cross sections~\cite{Ball:2011mu}. These correlations are shown
in Figs.~\ref{fig:correlations1} and ~\ref{fig:correlations2} for various configurations of the $\gamma$-jet
production at the LHC and for two values of the photon transverse energy: $\ETg=27.5$~GeV and
$\ETg=300$~GeV respectively.  
Photons+jets at central rapidities (top panels) have a dominant sensitivity to $g(x,Q^2)$ 
around $x$~=~0.01 for low $\ETg$, and around $x$~=~0.1 for high $\ETg$.
At forward jet rapidities (bottom panels) isolated-$\gamma$ probe the gluon and light-quark 
densities for a wide range of values at medium and small-$x$ for small and moderate $\ETg$, 
while for the highest $\ETg$ the light quarks are probed at very large-$x$.

%%%%%%%%%%%%%%%%%%%%%%%%%%%%%%%%
%%%%%%%%%%%%%%%%%%%%%%%%%%%

%%%%%%%%%%%%%%%%%%%%%%%%%%%%%%%%
%%%%%%%%%%%%%%%%%%%%%%%%%%%

%%%%%%%%%%%%%%%%%%%%%%%%%%%%%%%%%%%%%%%%%%%%%%%%%%%%%%%%%%%%%%%%%%%%%%%%%%%%%%%%%%%%%%%%
\section{Results}
%\label{sec:results}

\subsection{Analysis of current LHC photon-jet data}

\begin{figure}[htbp!]
\centering
\epsfig{width=0.32\textwidth,figure=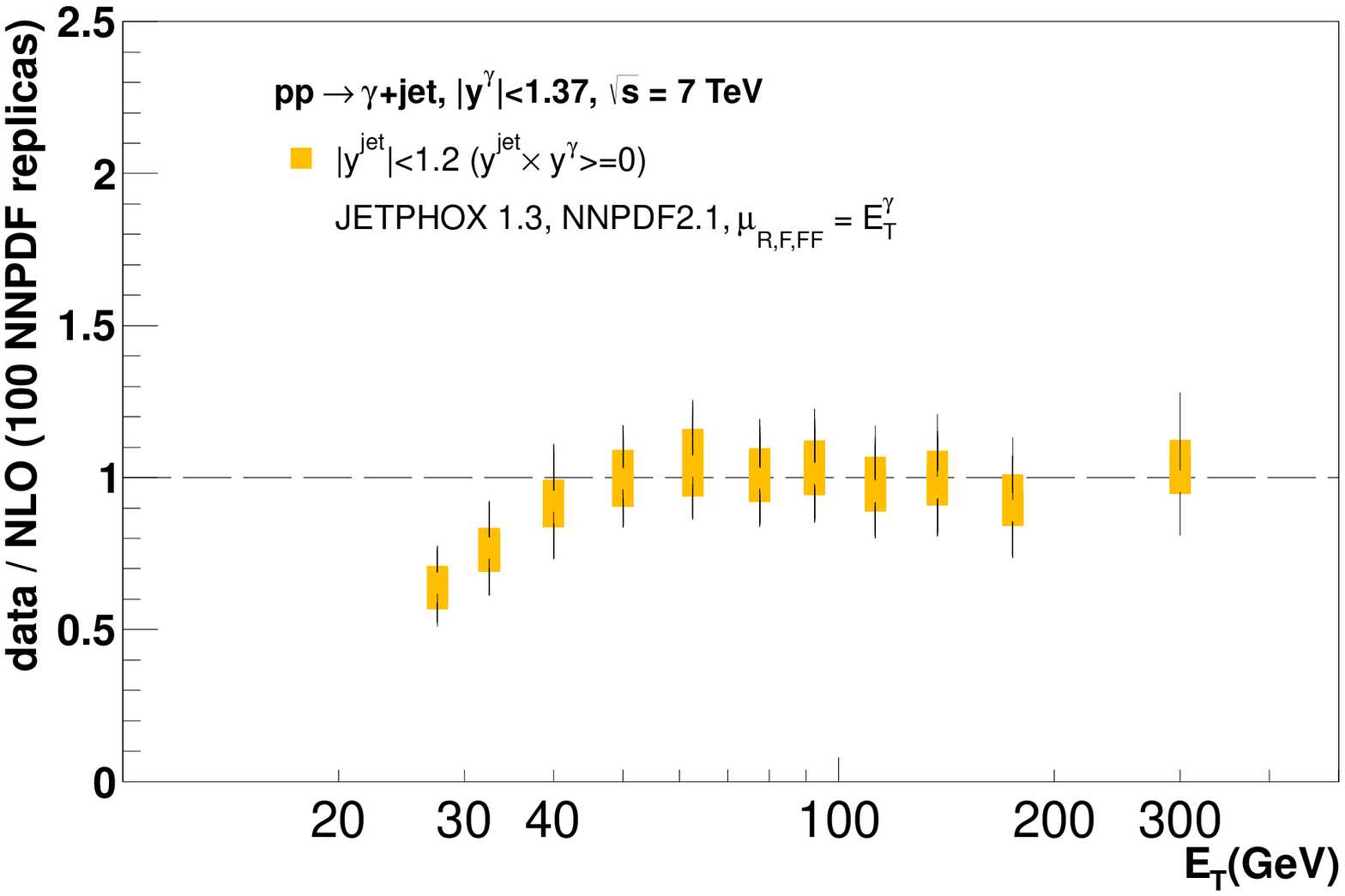}
\epsfig{width=0.32\textwidth,figure=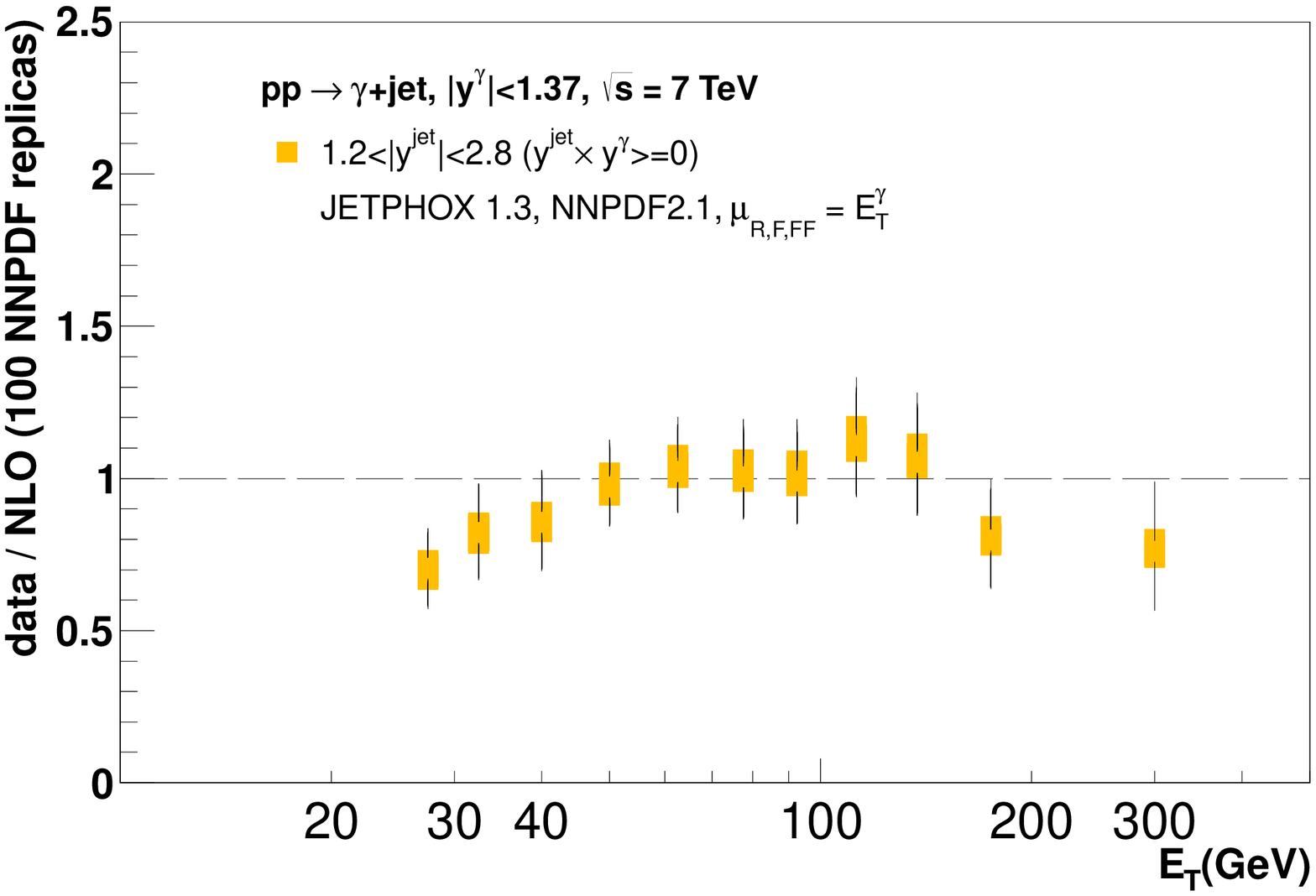}
\epsfig{width=0.32\textwidth,figure=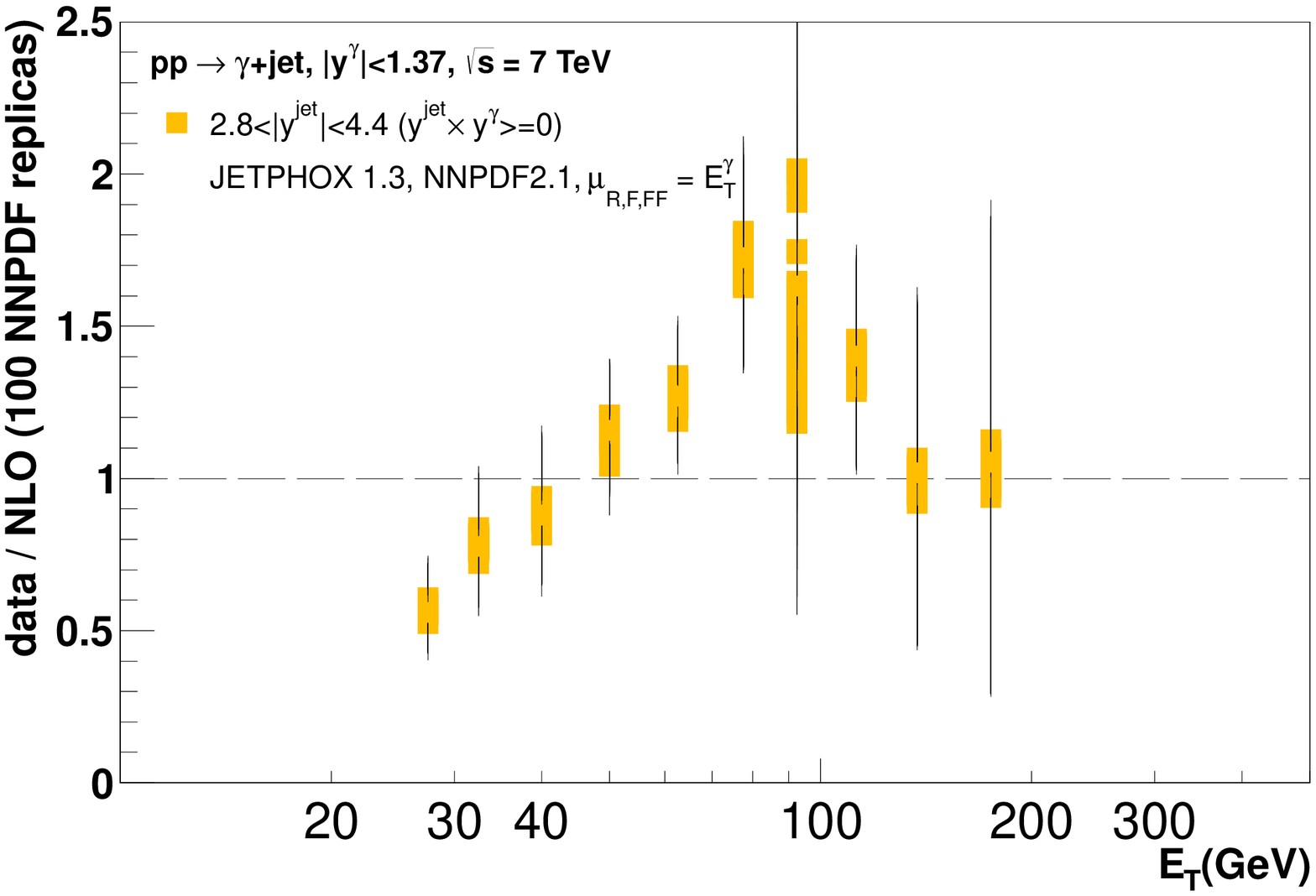}
\epsfig{width=0.32\textwidth,figure=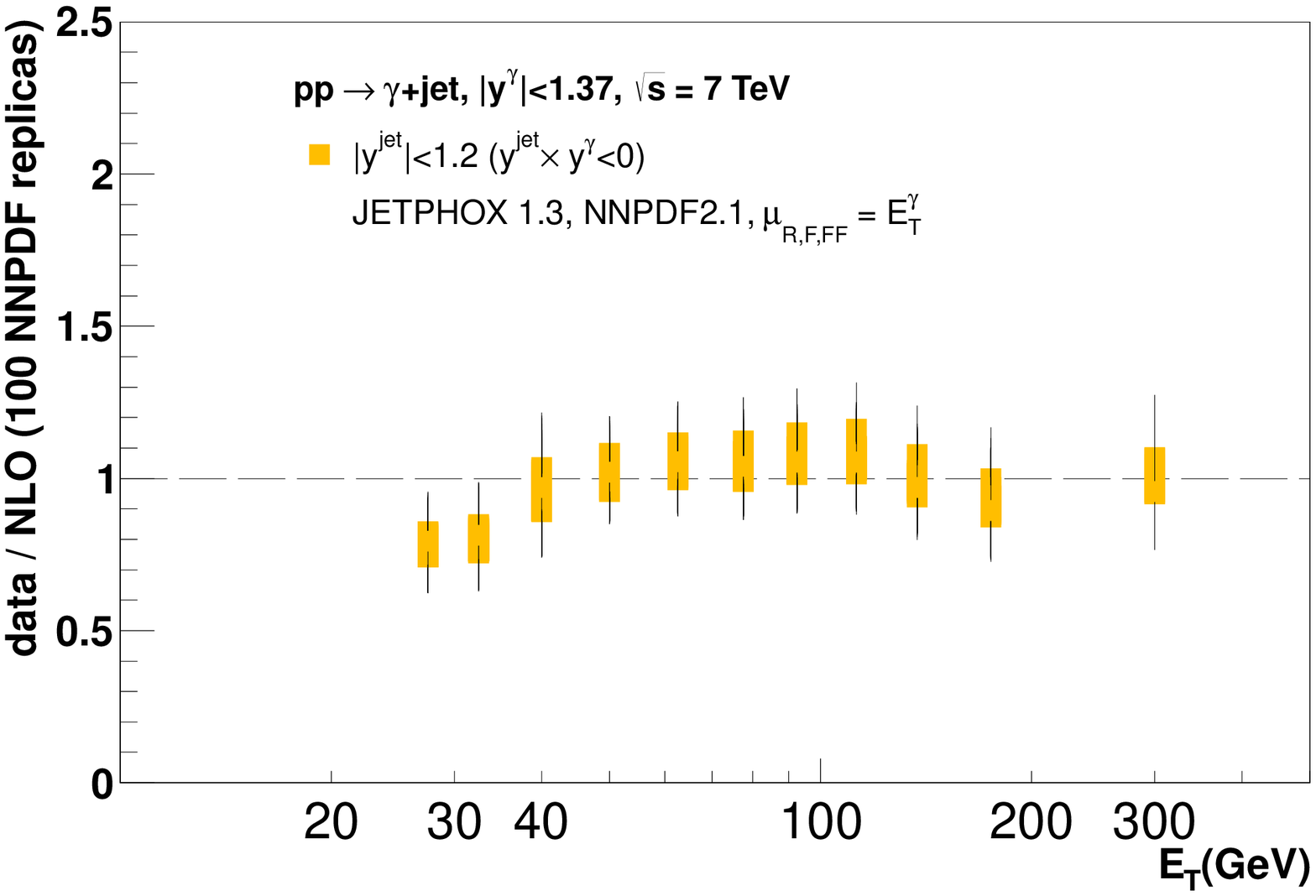}
\epsfig{width=0.32\textwidth,figure=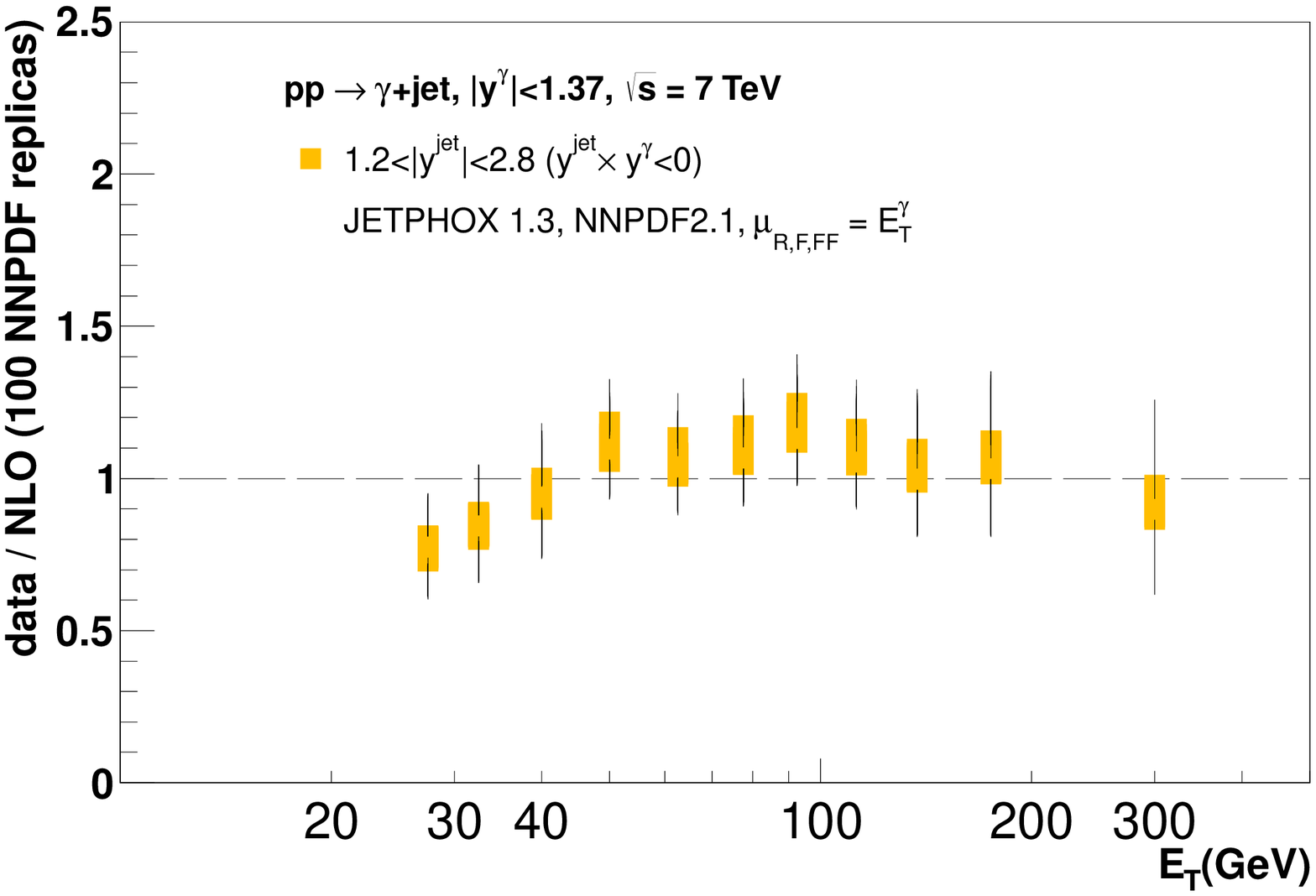}
\epsfig{width=0.32\textwidth,figure=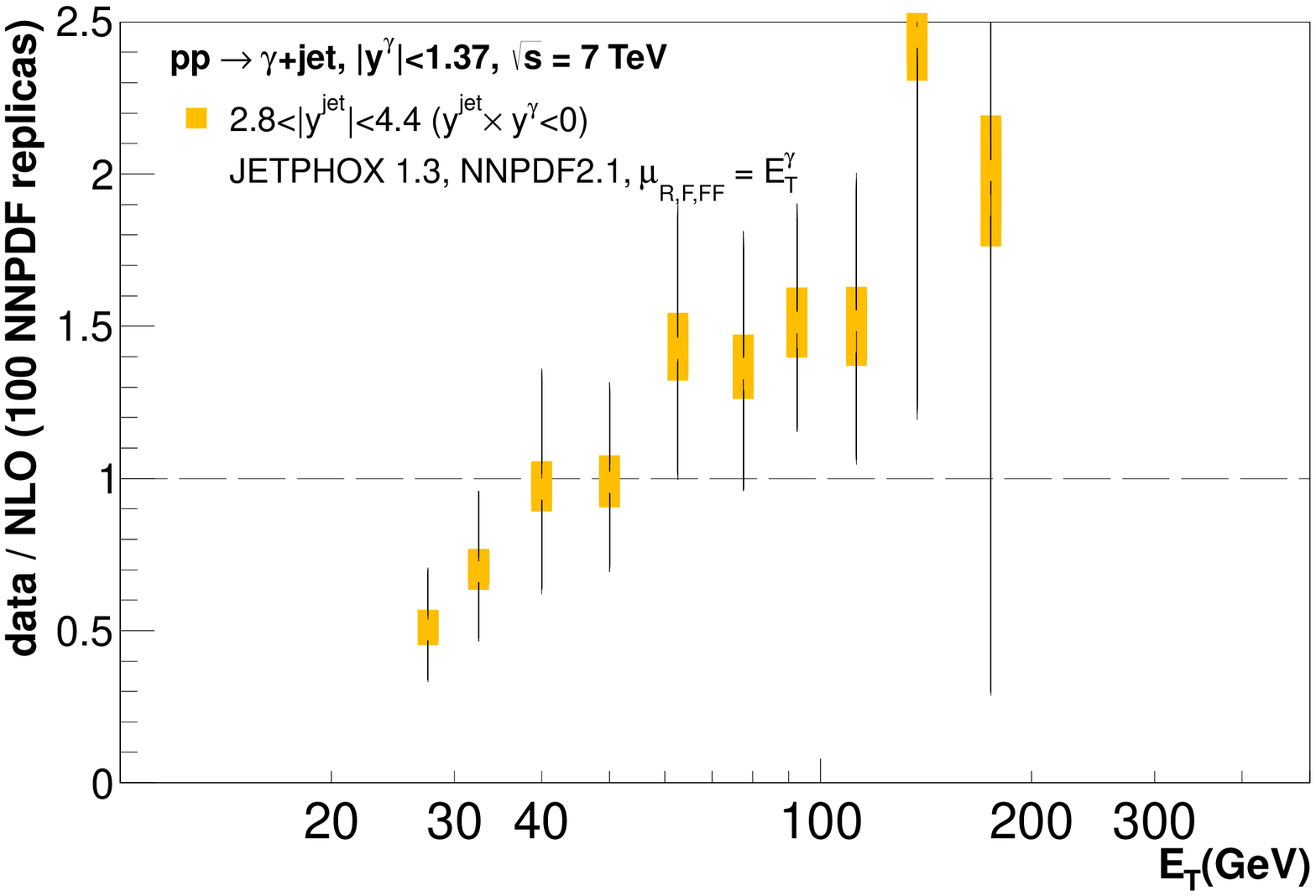}
\caption{\small Ratio of the ATLAS photon-jet $\ETg$-differential cross sections over NLO pQCD predictions in
  \pp\ collisions at $\sqrts$~=~7~TeV for the quoted jet rapidities. 
The top (bottom) panels show the case where photons and jets are produced in the same (opposite) hemispheres.
The (yellow) band indicates the range of predictions for each of the 100 NNPDF2.1 replicas, 
and the bars show the total experimental uncertainty.
\label{fig:datatheo_atlas}}
\end{figure}

\begin{figure}[hbp!]
\centering
\epsfig{width=0.65\textwidth,figure=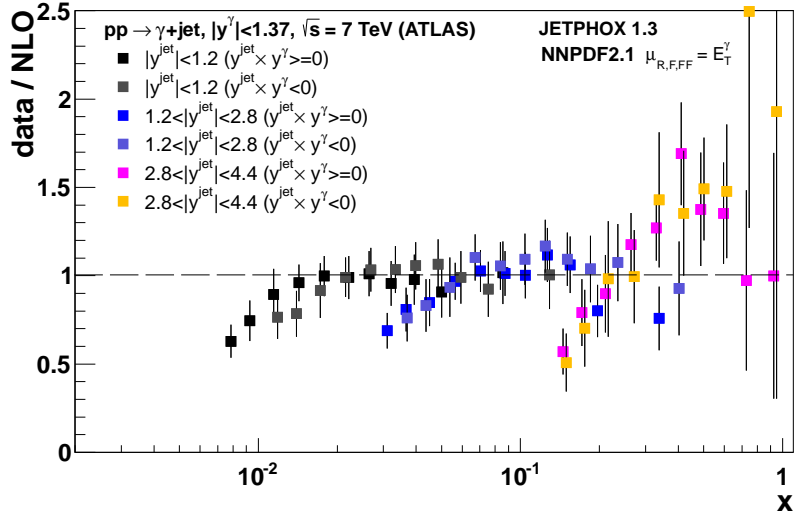}
\caption{\small Data/theory ratios for the six photon-jet distributions considered in this analysis, as a
function of $x = \ETg\,\big(e^{\mean{y^{\rm jet}}}-e^{-\mean{y^{\gamma}}}\big)/\sqrts$. 
For each system, the NLO prediction used is the one obtained with the central NNPDF2.1 replica.
The error bars indicate the total experimental uncertainty.
\label{fig:datatheo_summary}}
\end{figure}

Figure~\ref{fig:datatheo_atlas} shows the ratios between the experimental and theoretical  $d\sigma/d\ETg$
differential cross sections for the six jet rapidity ranges of the ATLAS measurement. The (yellow) band gives the
range of the predictions obtained for each of the 100 replicas, while the outer error-bars
cover the sum in quadrature of the statistical and (asymmetric) systematic uncertainties of the measurement. 
We observe an overall good agreement for all rapidity ranges within uncertainties, except  
for the most forward jets ($2.8 \leq |y^{\rm jet}| < 4.4$) where the data have a more concave shape than
the theory, and in the lowest bin ($\ETg$~=~27~GeV) where the theoretical predictions overshoots the central
value of the data points (a trend already observed in a few of the LHC inclusive isolated-$\gamma$
spectra~\cite{d'Enterria:2012yj}). Figure~\ref{fig:datatheo_summary} shows the comparison of the measured photon-jet
distributions with the NLO predictions obtained using the central NNPDF2.1 replicas, as a function of
$x = \ETg\,\big(e^{\mean{y^{\rm jet}}}-e^{-\mean{y^{\gamma}}}\big)/\sqrts$. 
The six data/theory ratios are around unity indicating an overall good agreement 
between NLO pQCD and the experimental measurements over the range $x\approx 10^{-2}-1$.

%%%%%%%%%%%%%%%
%%%%%%%%%%%%%%%%

%%%%%%%%%%%%%%%
%%%%%%%%%%%%%%%%

%%%%%%%%%%%%%%%%%%%%%%%%%%%%%%%%%%%%%%%%%%%%%%%%%%%%%%%%%%%%%%%%%%%%%%%%%%%%%%%%%%%%%%%%
%\subsection{Impact of isolated-photon data on the gluon}

Table~\ref{tab:chi2datasets} summarizes the quantitative results of our study. The first column lists the
average $\chi^2$ between each of the datasets and the NLO calculations over all replicas. For a 
majority of cases the agreement is quite good 
($\chi^2\approx$~1), while for the most forward jet configurations the $\chi^2$ obtained is poorer
($\chi^2\approx$~1.9 and $\chi^2\approx$~2.7 in the $\gamma$-jet opposite and same-hemisphere cases respectively). 
This result confirms at the quantitative level that there is an overall good agreement  
between NLO pQCD and the experimental isolated-$\gamma$+jet spectra measured at the LHC, as found previously
for all the inclusive isolated-$\gamma$ distributions~\cite{d'Enterria:2012yj}.\\

%%%%%%%%%%%%%%%%%%%%%
%\begin{sidewaystable}[htbp]
\begin{table}[Hhtbp!]
\caption{\small Summary of the $\chi^2$-analysis between NLO pQCD and the ATLAS
$\gamma$-jet data. For each system we list the initial data--theory $\chi^2$ (and its associated standard deviation),
the $\chi^2_{\rm rw}$ obtained after including each corresponding dataset via PDF reweighting, 
the mean $\la \alpha \ra $ of the associated $\mathcal{P}(\alpha)$ distribution, and the effective final
number of replicas after reweighting.
\label{tab:chi2datasets}}
\vspace{0.3cm}
%\begin{small}
%\begin{footnotesize}
%\begin{center} 
\centering
\begin{tabular}{cc|c|c|c|c}\hline
Rapidities & $\gamma$-jet $y$-hemisphere & \hspace{0.2cm} $\chi^2$ $\pm$ $\sigma_{\chi^2}$\hspace{0.2cm} & \hspace{0.2cm} $\chi^2_{\rm rw}$ \hspace{0.2cm} & \hspace{0.2cm} $\la \alpha \ra$ \hspace{0.2cm}  & \hspace{0.2cm} $N_{\rm eff}$ \hspace{0.2cm}  \\
\hline
$|y^{\rm jet}| < 1.2$, $|y^\gamma|<1.37$ & same & 1.1 $\pm$ 0.2  & 1.2 & 1.2 & 98 \\
$|y^{\rm jet}| < 1.2$, $|y^\gamma|<1.37$ & opposite & 0.7 $\pm$ 0.3 & 0.5 & 0.8 & 96 \\
$|y^{\rm jet}| = 1.2 - 2.8$, $|y^\gamma|<1.37$ & same & 1.3 $\pm$ 0.1 & 1.2 & 1.2 & 98 \\
$|y^{\rm jet}| = 1.2 - 2.8$, $|y^\gamma|<1.37$ & opposite & 0.7 $\pm$ 0.2  & 0.6 &0.9 & 99 \\
$|y^{\rm jet}| = 2.8 - 4.4$, $|y^\gamma|<1.37$ & same & 2.7 $\pm$ 0.6  & 2.6 & 1.9 & 81 \\
$|y^{\rm jet}| = 2.8 - 4.4$, $|y^\gamma|<1.37$ & opposite & 1.9  $\pm$ 0.3 & 1.9 & 1.5 & 96 \\
\hline 
\end{tabular}\vspace{3mm} 
%\end{small}
\end{table}
%%%%%%%%%%%%%%%%%%%%%%%%%%%%%%%%%%%%%%%

\begin{figure}[htpb!]
\centering
\epsfig{width=0.49\textwidth,figure=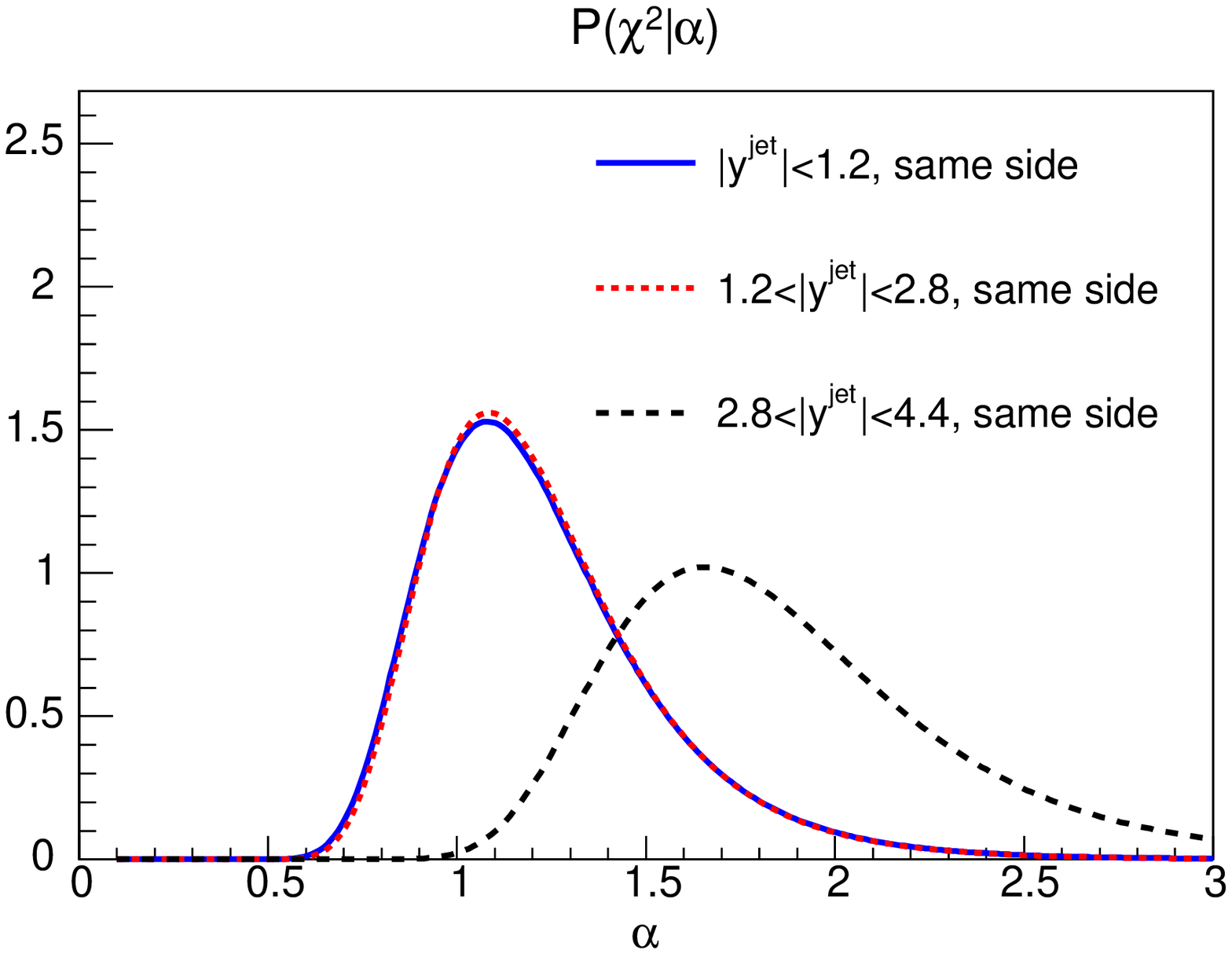}
\epsfig{width=0.49\textwidth,figure=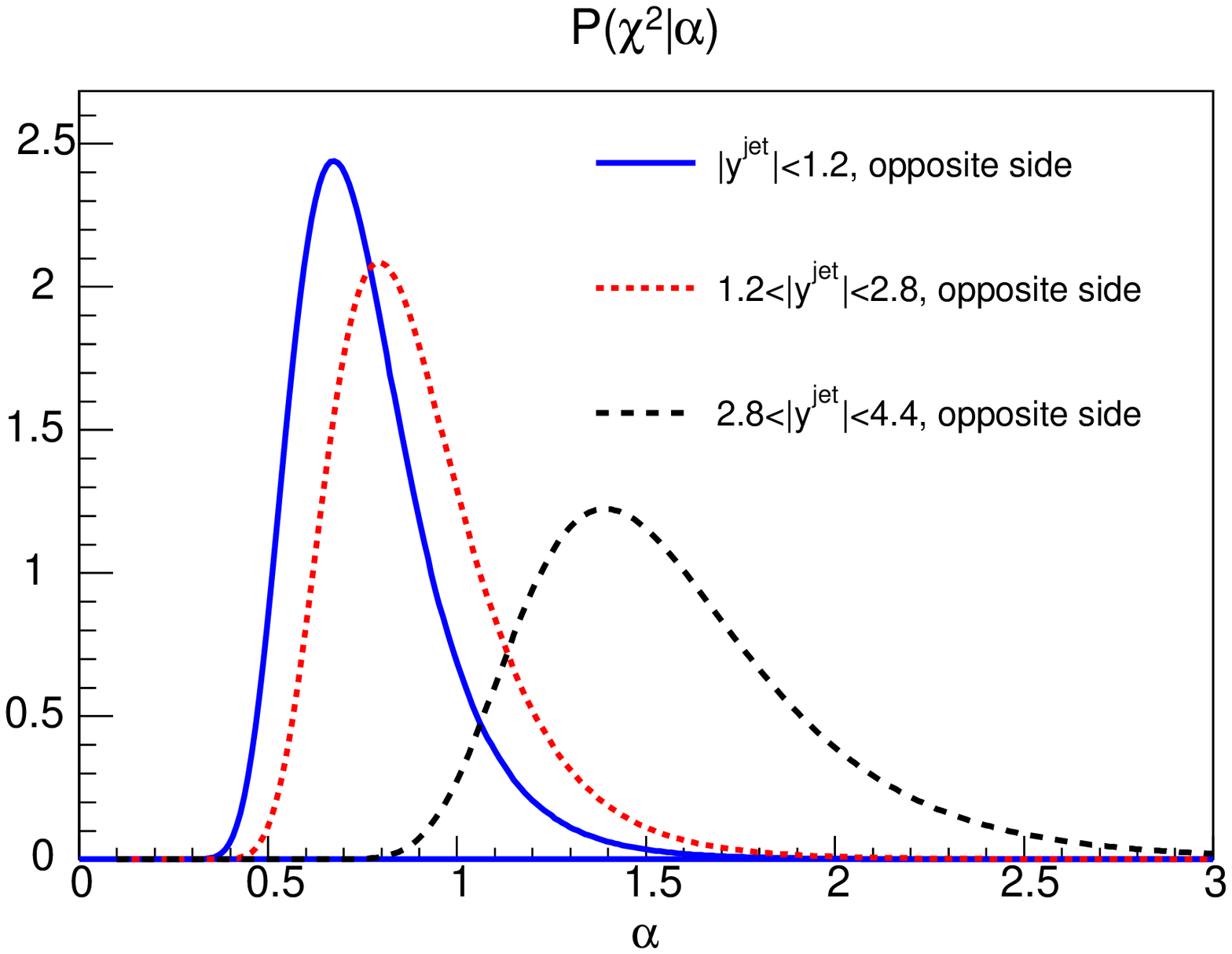}
\caption{\small Distribution of the $\alpha$ rescaling variable for the six jet rapidity bins 
of the ATLAS $\gamma$-jet data in \pp\ collisions at $\sqrts$~=~7~TeV.
The left (right) plot shows the results for the same (opposite) $\gamma$-jet $y$-hemisphere.
 \label{fig:palpha}}
\end{figure}

In Fig.~\ref{fig:palpha} we show the $\mathcal{P}(\alpha)$ distributions for the various kinematical configurations. 
In all cases $\mathcal{P}(\alpha)$ peaks close to one apart from the most forward jet data-samples, confirming
the overall consistency of these datasets with NLO pQCD and the proper estimation of the associated experimental errors.
%In most cases but the most forward data-samples, the distributions of uncertainties rescaling factors
%$\mathcal{P}(\alpha)$ peaks close to one. 
Since experimental uncertainties seem correctly determined for the measurement, the  $\la \alpha \ra\approx$~1.5--2
value of the most-forward jets results may point to larger theoretical uncertainties in this region 
due to some inadequacy of fixed-order NLO calculations for such a kinematical configuration. Similar differences
between data and theory for forward-central dijet distributions in \pp\ collisions have been observed
at 7~TeV~\cite{Chatrchyan:2012gwa}. 
Excluding the same-side large-rapidity data, the total initial $\chi^2$ 
of all the systems considered is $\chi^2=1.1$, while after 
reweighting it decreases to $\chi^2_{\rm rw}=1.0$. %for the whole dataset. 
The corresponding total effective number of replicas after reweighting is $N_{\rm eff}$~=~97.\\

%%%%%%%%%%%%%%%
%%%%%%%%%%%%%%%%

\begin{figure}[t]
\centering
\epsfig{width=0.49\textwidth,figure=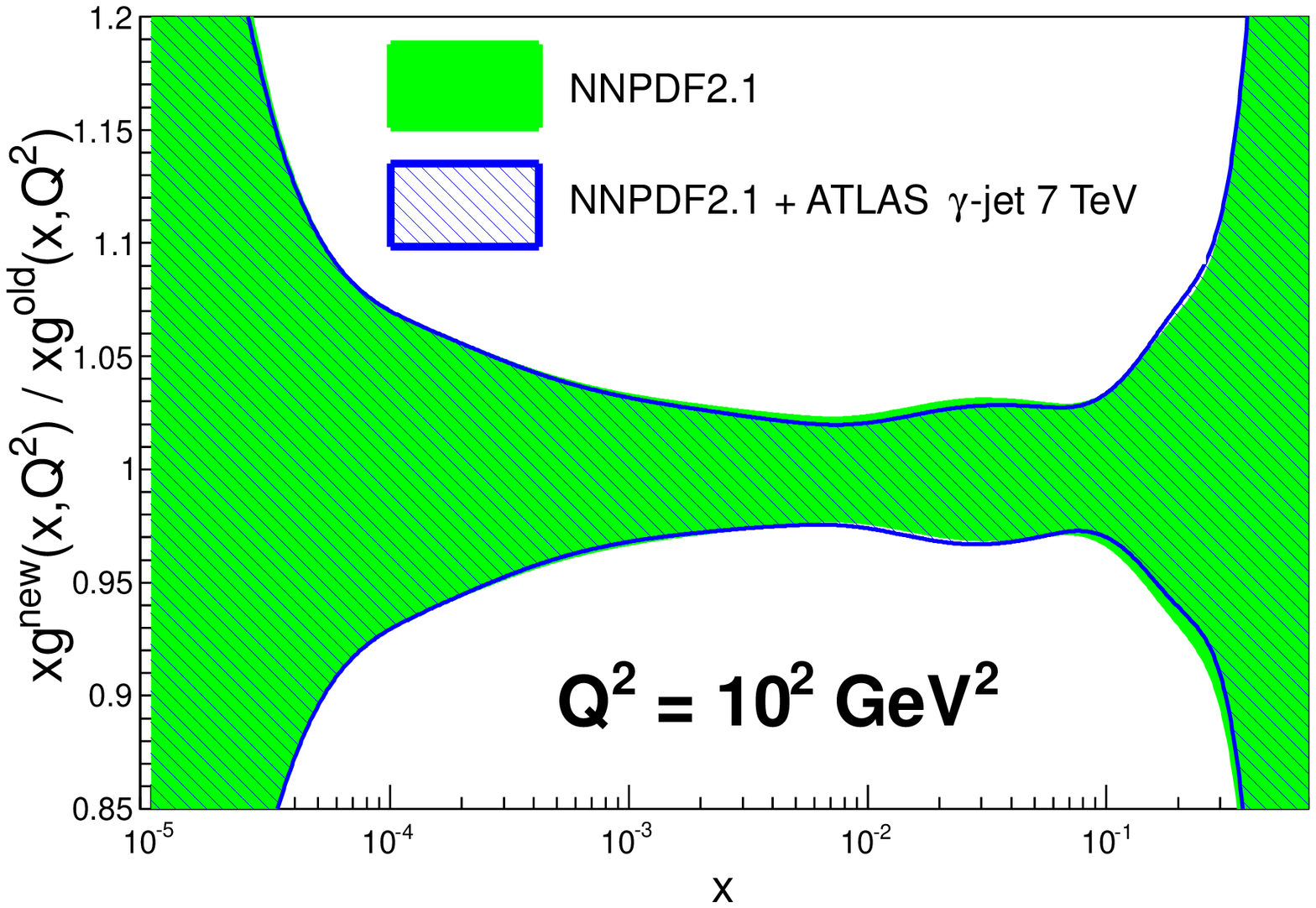}
\epsfig{width=0.49\textwidth,figure=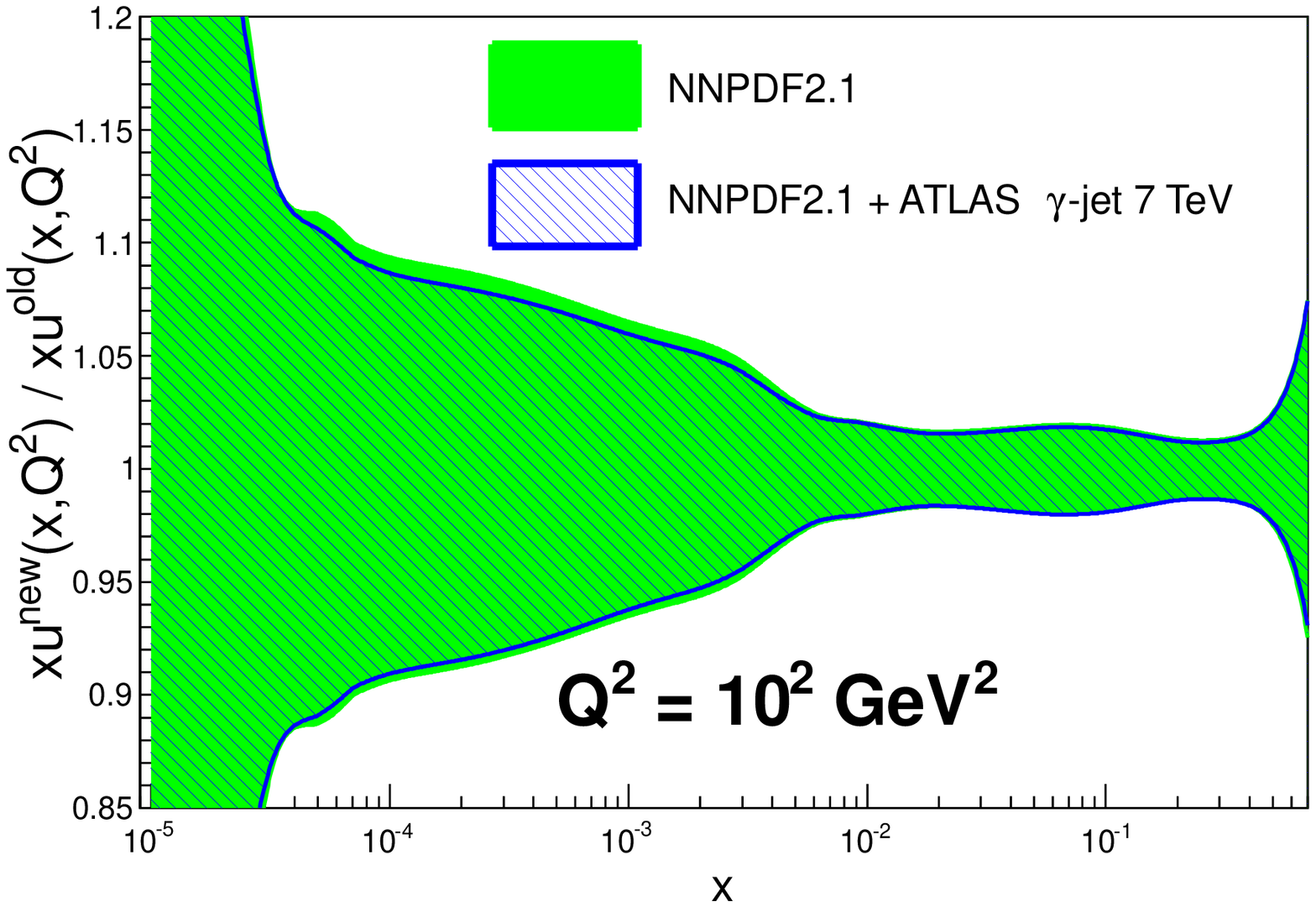}
\epsfig{width=0.49\textwidth,figure=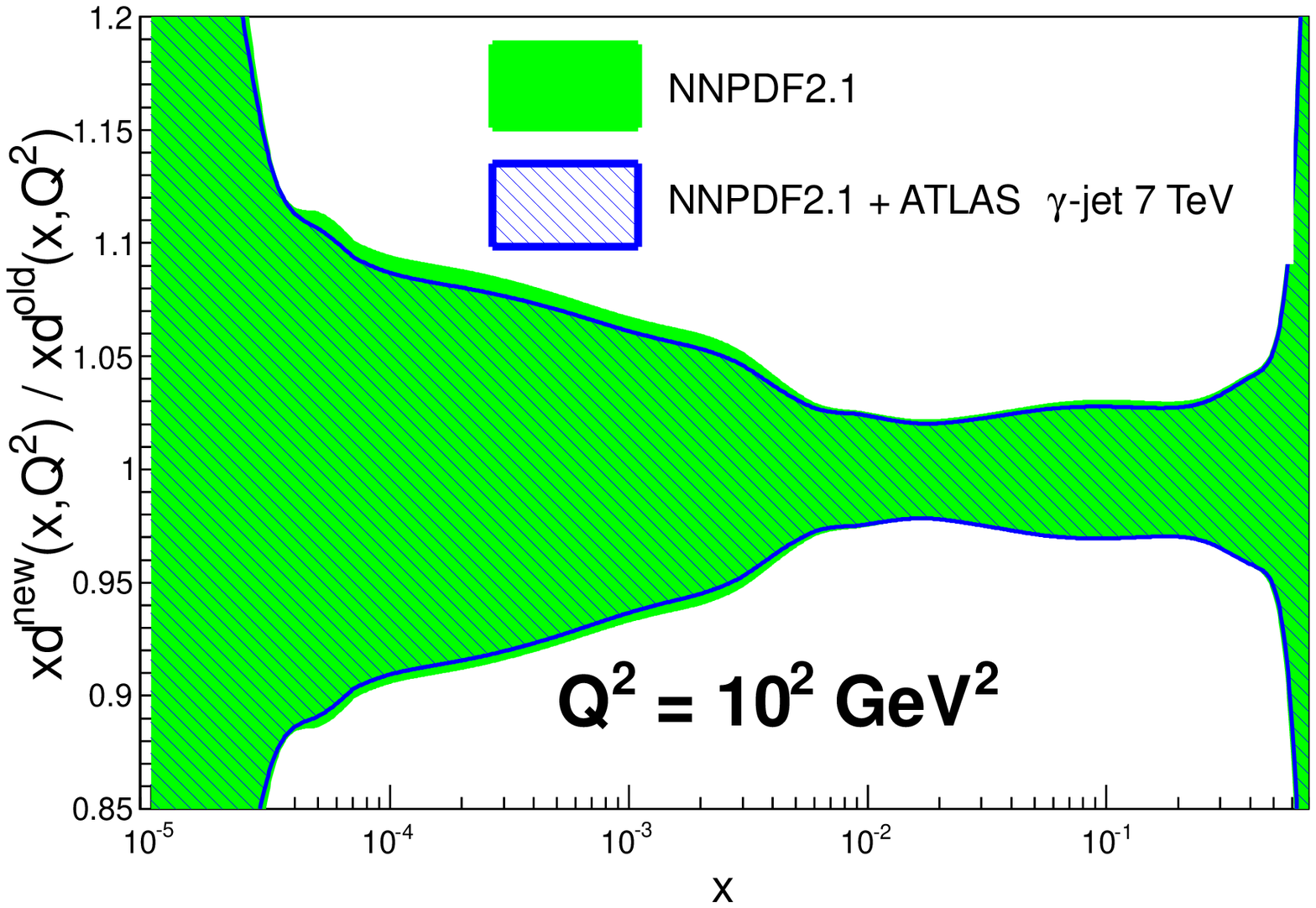}
\epsfig{width=0.49\textwidth,figure=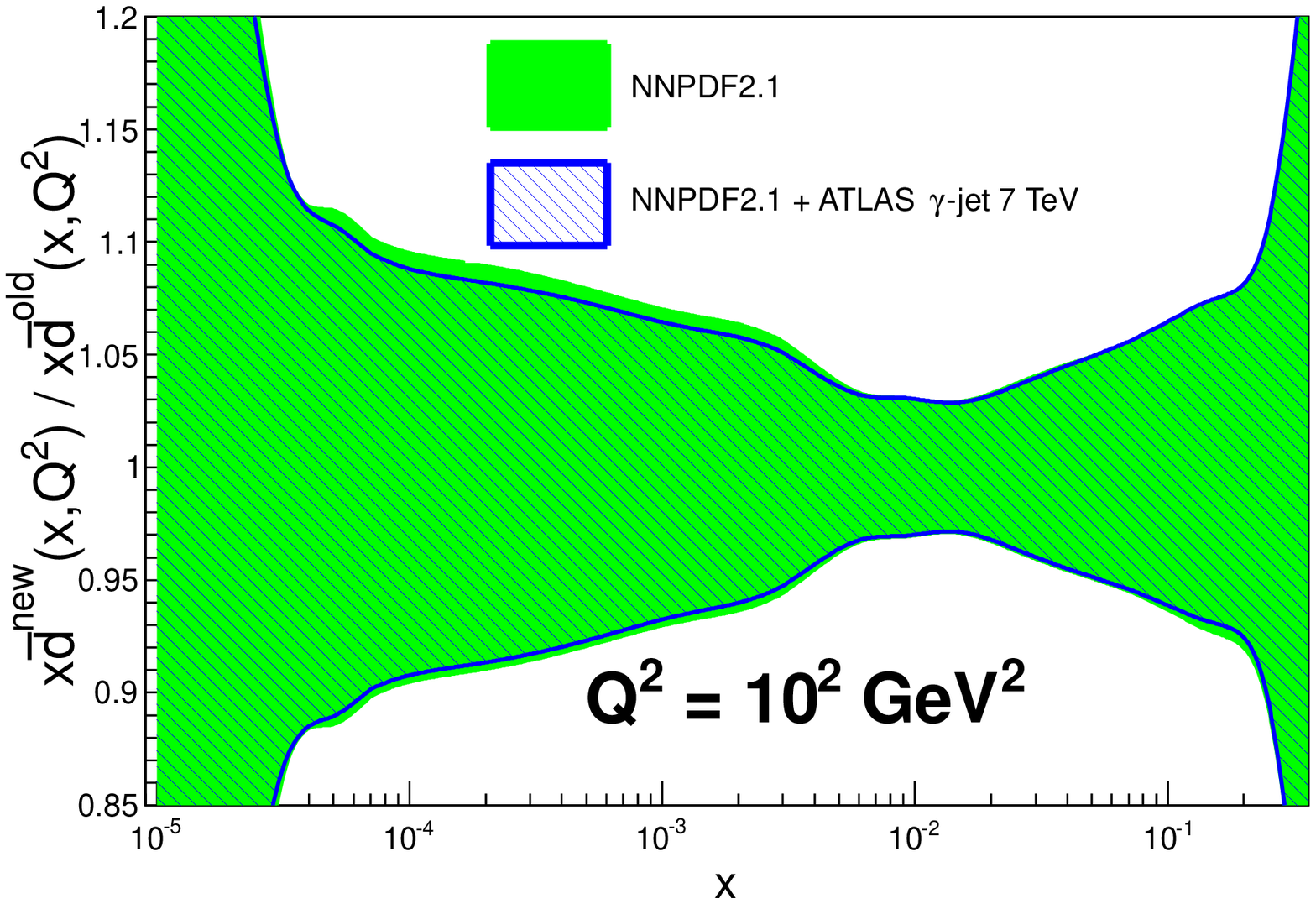}
\caption{\small Ratio between the NNPDF2.1 NLO PDF and associated uncertainties 
before (green solid band) and after (dashed blue area) inclusion of the ATLAS 
$\gamma$-jet data measured at 7~TeV. From top to bottom and from left to right 
we show the gluon, the up quark, the down quark and the anti-down quark.
 PDF are valued at $Q^2=100$~GeV$^2$.
\label{fig:gluon}}
\end{figure}

The direct quantification of the impact on the gluon and light-quark distributions is shown 
in Fig.~\ref{fig:gluon} where the ratios of NNPDF2.1 NLO PDF, evaluated at $Q^2$~=~100~GeV$^2$, are
plotted before and after including the ATLAS isolated-$\gamma$+jet data. 
%We also show the corresponding results for the light quark flavors.
The central values of the NLO parton densities are essentially unaffected by 
the new photon-jet LHC data which only lead to a rather mild (about 5\%) PDF 
uncertainty reduction at intermediate gluon fractional momenta $x\approx$~0.06 to 0.3 and
%. We also observe a similarly small improvement 
in the small-$x$ region between $10^{-4}$ and  $10^{-2}$ for light quarks.\\

%%%%%%%%%%%%%%%
%%%%%%%%%%%%%%%%

\subsection{Analysis of LHC photon-jet pseudodata with reduced uncertainties}
\label{subsec:pseudodata}

\begin{figure}[t]
\centering
\epsfig{width=0.49\textwidth,figure=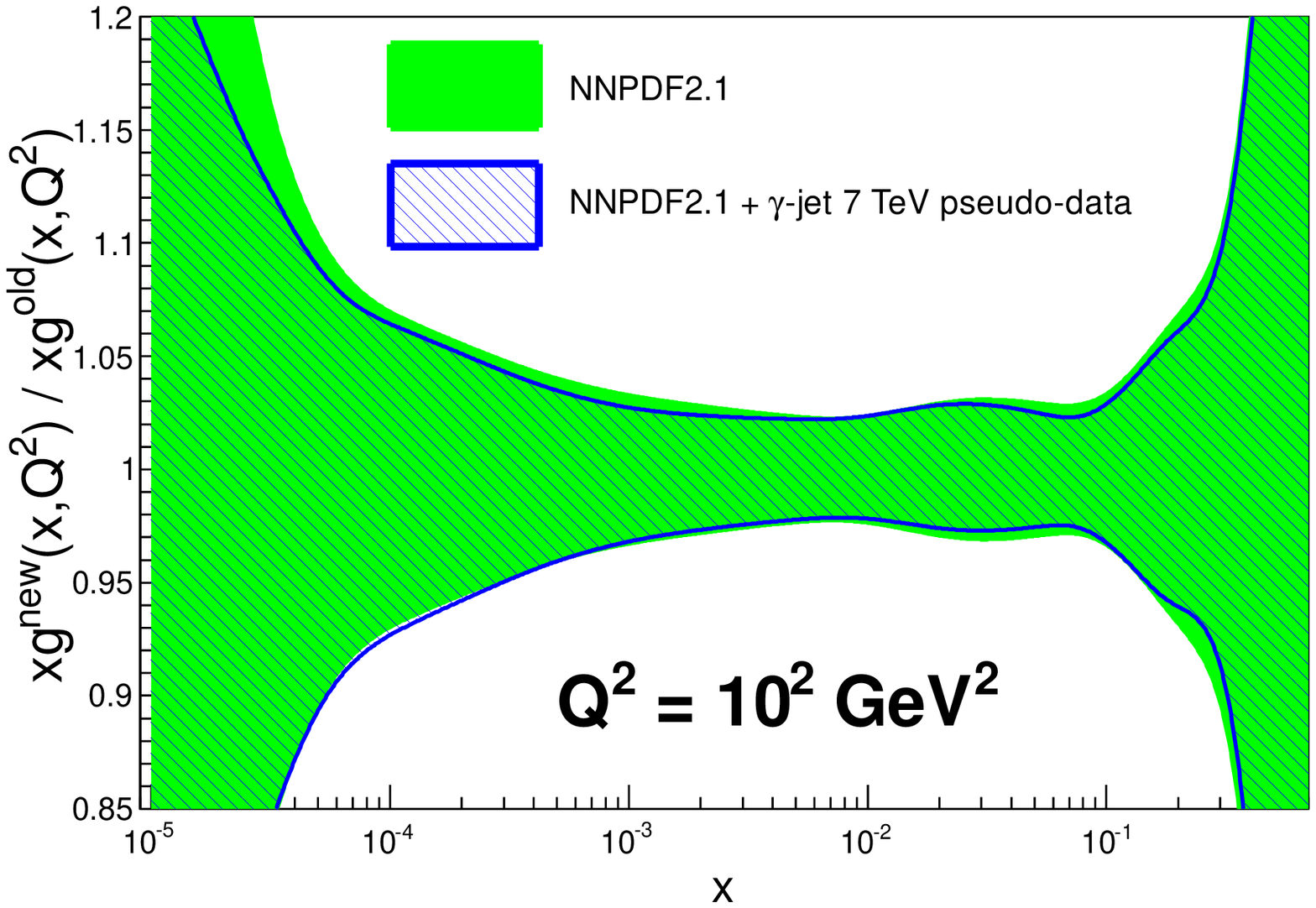}
\epsfig{width=0.49\textwidth,figure=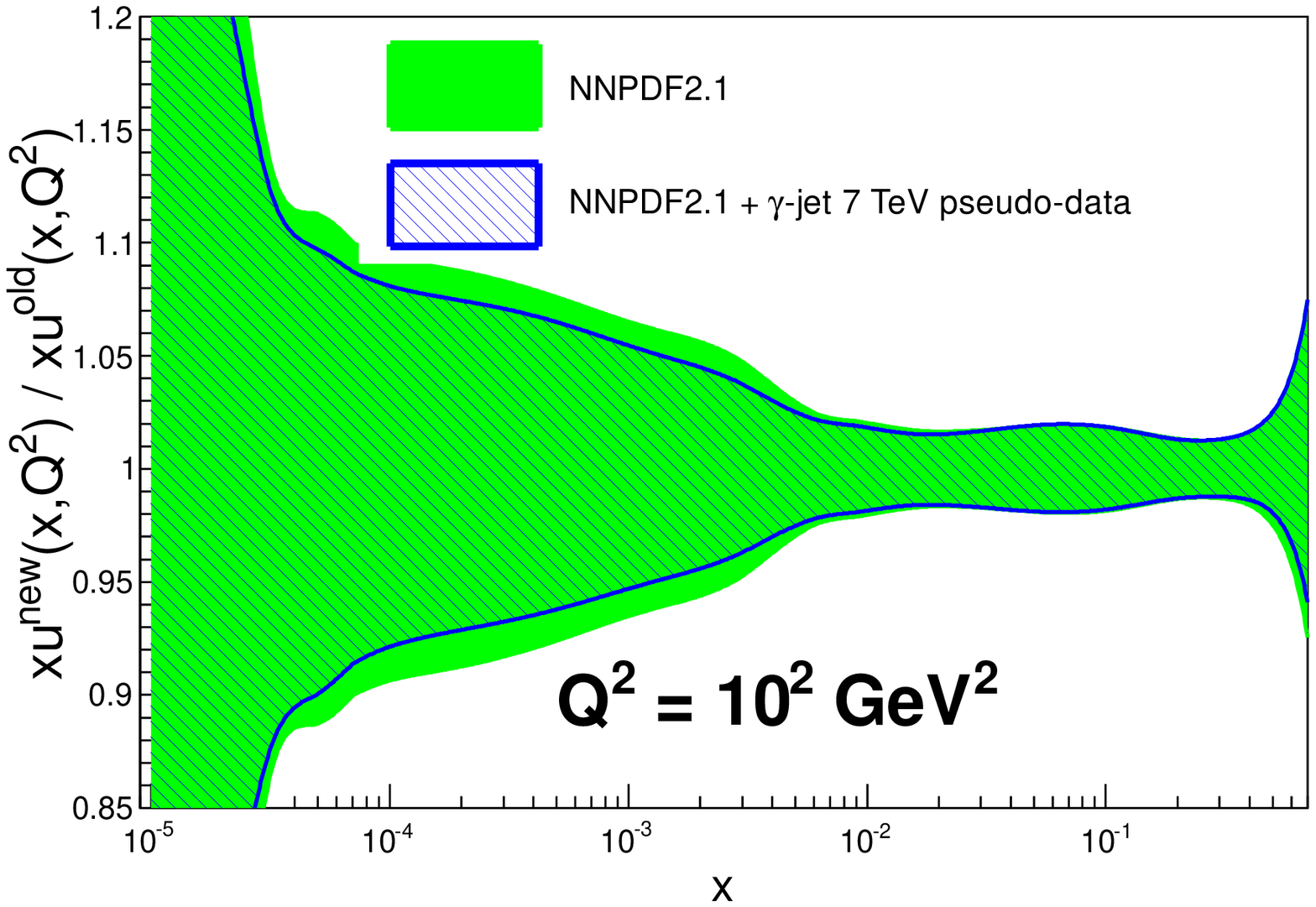}
\epsfig{width=0.49\textwidth,figure=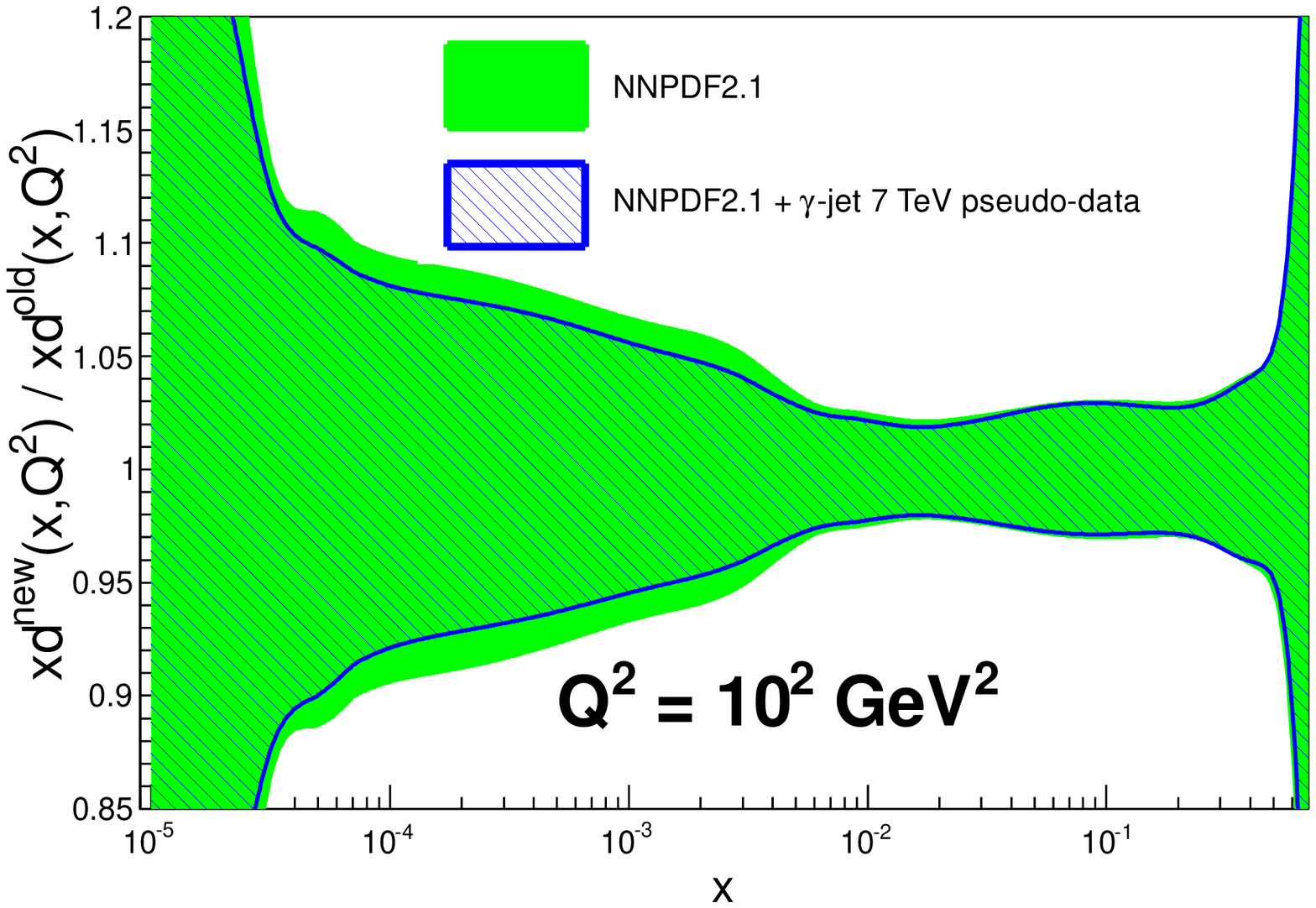}
\epsfig{width=0.49\textwidth,figure=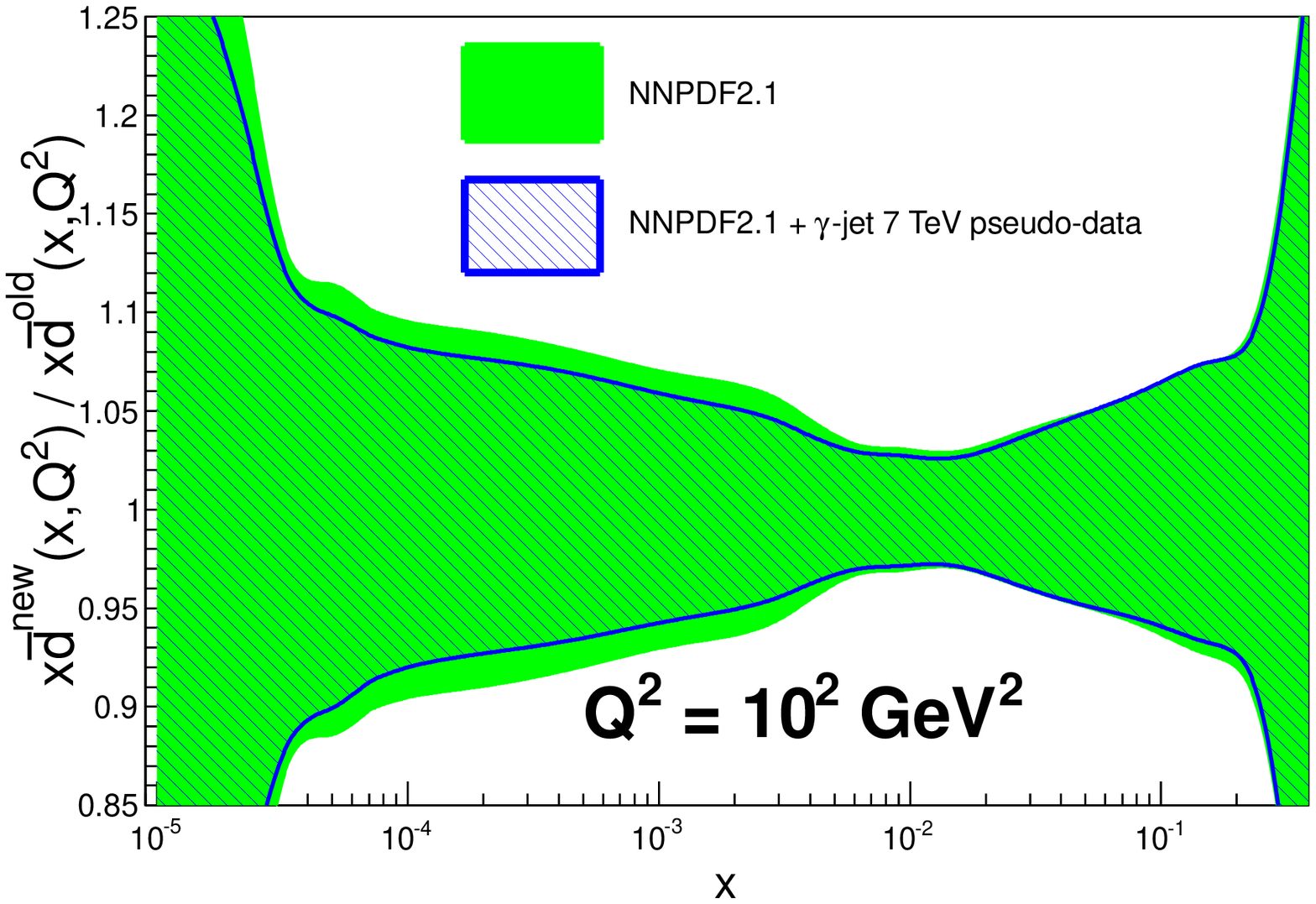}
\caption{\small
Same as Fig.~\ref{fig:gluon} in the case
of artificial  $\gamma$-jet pseudodata at 7~TeV assuming the same
  kinematical distributions of the existing ATLAS measurement
 but with reduced experimental uncertainties.
\label{fig:emulation}}
\end{figure}

The current relatively large experimental uncertainties of the available LHC photon-jet measurements,
as well as the lack of availability of their associated covariance error matrix,
result in a limited impact on the improvement of our knowledge of the proton PDF.
To quantify the possible sensitivity of future more precise LHC $\gamma$-jet data,
we have generated pseudodata for the same kinematics of the ATLAS measurement,
based on the NNPDF2.1 central predictions and assuming a total uncorrelated 
experimental uncertainty of $\pm$5\% ($\pm$10\%) for central and forward jets
and of $\pm$8\% ($\pm$15\%) for very-forward jets, above (below) $\ETg = 45$~GeV respectively.
Such a scenario represents a realistic improvement of about a factor of 2 with respect to the current
measurement. We generate the pseudodata adding  
Gaussian random fluctuations and carry out the same PDF reweighting analysis done
with real data. The resulting ratios of reweighted over current NLO PDF are shown 
in Fig.~\ref{fig:emulation} %the same PDF comparison as in Fig.~\ref{fig:gluon},
%which now gauge the impact of the 
for the simulated, more precise, $\gamma$-jet pseudodata.
We observe an improved sensitivity to the gluon and quark PDF in particular, 
for the latter, at small-$x$. The PDF uncertainties are reduced by up to 20\% 
in some $x$ regions. This result indicates that the inclusion of future 
differential photon-jet cross sections $d\sigma/d\ETg$ data into global PDF analyses 
has the potential to reduce the uncertainties of the gluon and light-quarks densities.

%%%%%%%%%%%%%%%
%%%%%%%%%%%%%%%%

%%%%%%%%%%%%%%%%%%%%%%%%%%%%%%%%%%%%%%%%%%%%%%%%%%%%%%%%%%%%%%%%%%%%%%%%%%%%%%%%%%%%%%%%
%\clearpage
\section{Summary }
%\section{Summary and outlook}
%\label{sec:summary}

We have quantified the impact on the proton PDF of the existing isolated-$\gamma$+jet $\ETg$-differential
cross sections in \pp\ collisions at 7~TeV where the photons are measured at central rapidities
($|y^\gamma|<1.37$) and the jets over $|y^{\rm jet}|<4.4$. Our theoretical setup includes NLO pQCD theoretical
calculations as implemented in the \jetphox\ program combined with the NNPDF2.1 parton densities and its associated PDF
reweighting technique. We find that NLO pQCD provides a good description of the photon-jet results at the LHC
in a wide kinematic range of photon transverse energies and jet rapidities, except maybe for the events 
where the jets are emitted at the most forward rapidities. The systematic uncertainties of the available measurements
are however still too large to provide significant constraints on the proton PDF. Nonetheless, our quantitative
studies with pseudodata confirm that future $\gamma$-jet measurements with reduced uncertainties can indeed
provide constraints on both the gluon density over a large $x$ domain as well as on the small-$x$ light-quarks
distributions. Photon-jet measurements at the LHC, as already shown in previous similar studies for inclusive
isolated prompt-$\gamma$, constitute thus an interesting ingredient of future global
PDF fits, complementary to the other data-sets currently used.

%In this work we have not considered the impact of the theory uncertainties
%in the NLO QCD cross sections. These uncertainties are large and
%might affect the potential of these data for PDF constraints, at least
%until the full NNLO calculation is not available. One possibility to 
%bypass this problem would be to use the cross section ratios between photon-jet cross sections between
%different LHC energies, since theory uncertainties cancel to a good extent leaving but
%the sensitivity to PDFs~\cite{Mangano:2012mh}.

\vspace{0.5cm}
{\bf\noindent  Acknowledgments \\}
The research of J.~R. has been supported by a Marie Curie Intra--European Fellowship of the European 
Community's 7th Framework Programme under contract number PIEF-GA-2010-272515.

\bigskip

%%%%%%%%%%%%%%%%%%%%%%%%%%%%%%%%%%%%%%%%%%%%%%%%%%%%%%%%%%%%%%%%%%%%%%%%%%%%%%%%%%%%%%%%
\clearpage

%\bibliography{photonsNNPDF}

\end{document}